\documentclass[11pt,a4paper]{article}
\usepackage{float}
\usepackage{graphicx}
\usepackage{amsmath}
\usepackage{amssymb}
\usepackage{latexsym}
\usepackage{color}
\usepackage{scrextend}
\usepackage{lipsum}
\usepackage[T1]{fontenc}
\usepackage{hyperref}
\usepackage{footnote}
\usepackage[utf8]{inputenc}
\usepackage{remreset}
\usepackage[utf8]{inputenc}
\usepackage[bottom]{footmisc}
\usepackage{amsfonts}
\usepackage{amsmath}
\usepackage{amssymb}
\usepackage{float}
\usepackage{graphicx}
\usepackage{makecell}
\usepackage{subfigure}%
\usepackage{lmodern}
\usepackage{microtype}
\usepackage{amsmath}
\usepackage{blindtext}
\usepackage{hyperref}
\usepackage[bottom]{footmisc}
\usepackage{bm}

\usepackage{subfigure}%
\usepackage{lmodern}
\usepackage{microtype}
 \usepackage{blindtext}
\providecommand{\U}[1]{\protect \rule{.1in}{.1in}}
\providecommand{\U}[1]{\protect \rule{.1in}{.1in}}

\begin{document}
\thispagestyle{empty}
\begin{center}

\vspace{1.8cm}

 {\large {\bf  Complex magnetism of the two-dimensional antiferromagnetic Ge$_2$F:
from a N\'eel spin-texture to a potential antiferromagnetic skyrmion }}\\

\vspace{1.5cm}

{\bf Fatima Zahra Ramadan$^{1}$}, {\bf Flaviano Jos\'e dos Santos$^{2}$}, {\bf  Lalla Btissam Drissi$^{1,2,3}$}, {\bf Samir Lounis$^{2,4}$}

\vspace{0.5cm}

$^{1}${\it LPHE-Modeling and Simulation, Faculty  of Sciences,
University
Mohammed V,\\ Rabat, Morocco}\\[1em]

$^{2}${\it Peter Gr\"{u}nberg Institut and Institute of Advanced Simulation,
Forschungszentrum J\"{u}lich \& JARA, D-52428, J\"{u}lich, Germany}\\[1em]

$^{3}${\it CPM, Centre of Physics and Mathematics, Faculty of Science, Mohammed V University in Rabat, Rabat, Morocco}\\[1em]

$^{4}${\it Faculty of Physics, University of Duisburg-Essen, 47053 Duisburg, Germany}\\[1em]

\vspace{3cm} {\bf Abstract}
\end{center}

Based on density functional theory combined with low-energy models, we
explore the magnetic properties of a hybrid atomic-thick two-dimensional
(2D) material made of Germanene doped with fluorine atoms in a
half-fluorinated configuration (Ge$_2$F). The Fluorine atoms are highly
electronegative, which induce magnetism and break inversion symmetry,
triggering thereby a finite and strong Dzyaloshinskii-Moriya interaction
(DMI). The magnetic exchange interactions is of antiferromagnetic nature
among the first, second and third neighbors, which leads to magnetic
frustration. The N\'eel state is found to be the most stable state, with
magnetic moments lying in the surface plane. This results from the
out-of-plane component of the DMI vector, which seems to induce an effective
in-plane magnetic anisotropy. Upon application of a magnetic field,
spin-spirals and antiferromagnetic skyrmions can be stabilized. We
conjecture that this can be realized via magnetic exchange fields induced by
a magnetic substrate. To complete our characterization, we computed the
spin-wave excitations and the resulting spectra, which could be probed via
electron energy loss spectroscopy, magneto-Raman spectroscopy or scanning
tunneling spectroscopy.
 \newline

\textbf{Keywords} {\begin{small}
Germanene; first principles calculations; Spin-wave; Skyrmions; Dzyaloshinskii-Moriya interaction (DMI); Wannier functions.
\end{small}}

\newpage
\section{Introduction}
The realization of complex spin-textures hinges on the presence of competing
magnetic interactions, which are heavily explored in various materials.
While hosting fundamentally exciting phenomena, such magnetic states have
great potential in spintronics with possible impact on information
technology. For example, going beyond the conventional ferromagnetic (FM)
materials for practical applications, such as the antiferromagnetic (AFM)
ones~\cite{jungwirth_antiferromagnetic_2016,Wadley2016,Olejnik2018,Baltz2018}
can have various interesting advantages. AFMs are expected to be robust
against perturbation due to magnetic fields, they produce no stray fields,
display ultrafast dynamics, and are capable of generating large
magnetotransport effects \cite{Baltz2018}.

The emergence of complex magnetic states is favored by the presence of
competing magnetic interactions, which can lead to frustration. Spin-orbit
driven interactions, such as the Dzyaloshinskii-Moriya Interaction (DMI),
also favor non-collinearity with the additional injection of a potential
chiral magnetic behavior. Indeed, broken inversion symmetry and spin-orbit
coupling triggers a finite DMI, which stabilizes a unique sense of rotation
of the magnetic moments. Various chiral spin-swirling states can then be
produced, such as chiral spin spirals, chiral domain walls or magnetic
skyrmions. The latter are topological protected vortex lines in which the
spins point in all the directions wrapping a sphere~\cite%
{Bogdanov1989,Roessler2006,muhlbauer2009}, which are promising for potential
high-density and low-power spintronics technology \cite%
{jiang2015,du2015,zhou2014}. In this field, there is currently a great
interest in going beyond FM skyrmions by discovering AFM skyrmions\cite%
{barker_static_2016,Velkov2016,Keesman2016,Zhang2016a,Goebel2017b,Kravchuk2019,diaz_topological_2019,Liu2020}%
, which would combine the advantages of skyrmions\cite%
{nagaosa2013,wiesendanger2016,fert2017,kang2016,zhang2015b,crum2015,fernandes2020}
and AFM properties.

The goal of this manuscript is to prospect the presence of chiral complex
spin textures in two dimensional (2D) magnetic materials. They not only
offer unique physical and chemical properties, but also an unprecedented
flexibility in system design. When grown in a multilayer fashion, their
flexibility stems from van-der-Waals bonding between neighboring
atomic-thick layers of potentially very different properties, which permits
virtually unlimited combinations and stackings of individual layers. The
resulting properties, usually conveyed via proximity effects, can be very
distinct from the original building block materials. Most of 2D materials do
not exhibit DMIs because of their centrosymmetric crystal structure.

To break such a symmetry, some approaches consist on creating 2D structures
within which different atoms are mixed in an alternating manner to generate
one-atom thick hybrids \cite{kan2010,sahin2009,drissi2019}. Other strategies
such as applying a bias voltage or strain are also used \cite%
{yao2008,liu2019,bai2015}. Chemical functionalization, impurities,
boundaries and defects are other efficient ways employed in 2D sheets to
tune their physical properties and induce magnetic order \cite%
{zhu2018,drissi2018,sun2017,liu2019, drissi2017,mao2018}. In particular,
chemisorption using radicals such as oxygen, hydrogen or fluorine atoms on
the surface of 2D honneycomb structures leads to long-range magnetism \cite%
{kan2010,ma2011,le2016,marsusi2019,zhou2010}. Another example consists of Sn
monolayer on SiC(0001) surface, where a strong spin-orbit coupling was found
on the basis of a generalized Hubbard model. This mainly contributes in the
formation of a nanoskyrmion state at realistic magnetic fields and
temperatures \cite{badrtdinov2018}.

Half-functionalization is also a powerful and widely-used tool to tailor
spin and magnetic behavior in 2D materials \cite{zheng2012,zhang2016}. A
particularly interesting adatom is fluorine since it is the most
electronegative element of the periodic table. Half fluorination is an
exothermic adsorption that generates stable 2D hexagonal structures \cite%
{drissi2016}. In half-fluorinated graphene, C$_2$F, where F-atoms form
strong covalent bonds with carbon, a threshold of the
antiferromagnetic-ferromagnetic instability with strong
Dzyaloshinskii-Moriya interaction was predicted \cite{mazurenko2016} with
the potential presence of ferromagnetic skyrmions. The latter work
challenged the ab-initio results obtained by Rudenko et al.~\cite%
{rudenko2013}, which revealed finite AFM interactions on the triangular
lattice of magnetic moments, leading to the instability of the collinear
magnetic ordering due to frustration and the stabilization of a $120^\circ$
N\'eel state. Mazurenko et al.~\cite{mazurenko2016} proposed that the direct
exchange interaction between spin orbitals, not accounted in Ref.~\cite%
{rudenko2013}, leads to a ferromagnetic interaction, which is capable of
compensating the antiferromagnetc indirect exchange interactions in C$_2$F.

Remarkably, half-fluorination can trigger opposite magnetic behavior in
hybrid 2D monolayers. While half-fluorinated BN sheet is an
antiferromagnetic direct semiconductor, half-fluoro-GaN monolayer shows
ferromagnetic character \cite{ma20117}. In silicene-graphene (SiC),
interesting magnetic properties can emerge depending depending on which host
atom (C or Si) fluorine is attached~\cite{drissi2013}.

In this paper, we study the presence of chiral spin-textures in half
fluorinated germanene using density functional theory (DFT) combined with
low-energy models with spin-orbit coupling in the spirit of the methodolgy
followed by Mazurenko et al.~\cite{mazurenko2016}. We found that the Ge$_{2}$%
F is antiferromagnetic with strong DMI between the first nearest magnetic
germanium neighbors. The spin dynamics simulations demonstrate the stability
of the antiferromagnetic N\'{e}el state, resulting from magnetic
frustration. In the latter configuration, the magnetic moments lie in the
surface plane, which is induced by the out-of-plane component of the DMI
vector. Extremely large magnetic fields can stabilize an antiferromagnetic
skyrmion. We conjecture that this can be enabled by a proximity-effect
induced by an underlying magnetic substrate. Noting that magnons in 2D
structures have been probed with magneto-Raman spectroscopy~\cite{cenker2020}
and scanning tunneling microscopy~\cite{klein1218}, we finally explore the
spin-wave excitations characterizing the obtained complex spin-textures.

\section{Computational details}

The electronic and magnetic properties have been obtained using the Quantum
espresso code~\cite{giannozzi2009}, which is based on density functional
theory (DFT). Exchange and correlation effects were taken into account using
the local spin density approximation (LDA). Spin-orbit (SO) coupling was
included on the basis of fully relativistic pseudopotentials. In these
calculations, we set the energy cutoff to $60 Ry$ for the plane-wave basis.
For the Brillouin-zone integration a $30\times 30\times 1$ Monkhorst Pack
mesh was used. To avoid the artificial interactions between layers, the
thickness of the vacuum space was fixed at $20\mathring{A}$.
To extract the magnetic exchange interactions required for the exploration
of the magnetic properties, we built a low-energy model using an effective
Hamiltonian following the work of Mazurenko et al.~\cite{mazurenko2016}: 
\begin{equation}
\hat{H}=\sum_{ij,\sigma \sigma ^{\prime }}t_{ij}^{\sigma \sigma ^{\prime }}%
\hat{a}_{i\sigma }^{+}\hat{a}_{j\sigma ^{\prime }}+\dfrac{1}{2}%
\sum_{i,\sigma \sigma ^{\prime }}U_{00}\hat{a}_{i\sigma }^{+}\hat{a}%
_{i\sigma ^{\prime }}^{+}\hat{a}_{i\sigma ^{\prime }}\hat{a}_{i\sigma }+%
\dfrac{1}{2}\sum_{i,\sigma \sigma ^{\prime }}U_{ij}\hat{a}_{i\sigma }^{+}%
\hat{a}_{i\sigma ^{\prime }}^{+}\hat{a}_{i\sigma ^{\prime }}\hat{a}_{i\sigma
}+\dfrac{1}{2}\sum_{i,\sigma \sigma ^{\prime }}J_{ij}^{F}\hat{a}_{i\sigma
}^{+}\hat{a}_{i\sigma ^{\prime }}^{+}\hat{a}_{i\sigma ^{\prime }}\hat{a}%
_{i\sigma }  \label{aa}
\end{equation}

where $i(j)$ and $\sigma (\sigma ^{\prime })$ are site and spin indices, $%
\hat{a}_{i\sigma }^{+}$ ($\hat{a}_{j\sigma ^{\prime }}$) are the creation
(annihilation) operators, and $U_{00}$, $U_{ij}$ and $J_{ij}^{F}$ represent
local Coulomb, non-local Coulomb and non-local ($i\ne j$) exchange
interactions, respectively, and are obtained using the constrained random
phase approximation (cRPA) \cite{amadon2014} as implemented in the ABINIT
code \cite{gonze2016}. $t_{ij}$ is a hopping matrix-element taking into
account the spin-orbit coupling, which is determined using the Wannier
parameterization for the three nearest neighbours. To parameterize the
(DFT+SO) spectra and construct the corresponding low-energy model, we use
maximally localized Wannier functions, implemented in the wannier90 package 
\cite{mostofi2008}.

Moreover, we used the Spirit atomistic spin dynamics simulation code \cite%
{muller2019} to solve the Landau-Lifshitz-Gilbert (LLG) equation: 
\begin{equation}
\frac{d\mathbf{M}_i}{dt}=-\gamma \mathbf{M}_i\times \mathbf{B}^\text{eff}_i+%
\frac{\alpha }{M_i} \mathbf{M}_i \times \frac{\partial \mathbf{M}_i}{%
\partial t}  \label{LLG}
\end{equation}
where $\gamma $ is the electronic gyromagnetic ratio, $\alpha $ the damping
factor, and $\mathbf{M}_i$ is the magnetic moment at a given site $i$. This
permits the investigation of the magnetic properties of Ge$_2$F described by
an extended Heisenberg Hamiltonian given in eq.~\ref{eq:heisenberg}, with $%
\mathbf{B}^\text{eff}_i = -\frac{\partial H}{\partial \mathbf{M}_i} $. We
assume a supercell of a size of $100\times 100\times 1$ atoms.

Once the ground state or a metastable state is found, we compute the
adiabatic spin-wave modes and the corresponding inelastic scattering
spectrum, based on time-dependent perturbation theory. The associated
theoretical framework was presented in Refs.\cite{dos_santos_first-principles_2017,dos_santos_spin-resolved_2018} and used
for various problems~\cite{dosSantos:2020a,dosSantos:2020b}. The spin-wave eigenvalues $\omega\mathbf{k}$ and eigenvectors $K$ are then
obtained after diagonalization of the system's dynamical matrix in the
reciprocal space. We arrive to the total dynamical structure factor (summing
up all the scattering channels), which is given by 
\begin{equation}
\Gamma(\mathbf{q},\omega) \propto \sum_{\alpha} \sum_{\mu\nu} e^{\mathrm{i} 
\mathbf{q} \cdot \mathbf{R}_{\mu\nu}} \mathcal{N}_{\mu \nu}^{\alpha \alpha} (%
\mathbf{q}, \omega) \quad ,
\end{equation}
where $\alpha,\beta = x,y,z$ and $\mu,\nu$ are site indices for spins in the
unit cell that encompasses the noncollinear ground state magnetic structure.
The spin-spin correlation tensor can be expressed using the information
about the spin-wave modes as 
\begin{equation}
\mathcal{N}_{\mu \nu}^{\alpha \beta} (q, \omega) = \sum_{k, r} \delta\left(\omega - \omega_{r}(k)\right) \langle 0\vert S_{\mu}^{\alpha(- q)}| k, r\rangle \langle k,r|S_{\nu}^\beta( q)|0\rangle ,
\end{equation}
where $\omega_r(\mathbf{k})$ is the energy of the spin-wave mode $r$ with
wavevector $\mathbf{k}$, and matrix elements of the spin operators between
the ground state and the excited spin-wave states~\cite%
{dos_santos_spin-resolved_2018}.

Within this framework we have access to several distinct scattering
channels. In this work we present results for the total inelastic scattering
spectrum due to spin waves (the sum over all scattering channels), as one
would measure in an experiment with an unpolarized scattering experiment
such as electron energy loss spectroscopy (EELS). The various scattering
channels will also be analyzed, which could be detected via the recent
theoretical proposal, spin-resolved EELS (SREELS), shown in Ref.~\cite%
{dos_santos_spin-resolved_2018}. Within the latter, a spin-polarized beam of
electrons is used to probe the magnetic material. The scattered electrons
are then spin-filtered with the spin analyzer collinear with the incident
beam polarization. This gives rise to four scattering channels, one for each
possible combinations of [incoming spin]-[outgoing spin]. Two of these
channels correspond to non-spin-flip processes, namely the up-up and the
down-down channels. The other two, up-down and down-up, account for
spin-flip events, where angular momentum is exchanged with the sample.

\section{Results and discussion}

In this paper, we study the magnetic properties of 2D half-fluorinated
germanene where F-atoms bind Ge-atoms occupying the A-sites of each
hexagonal lattice while B-sites remain undecorated as shown in Fig.~\ref%
{struct}. The bond lengths are $d_\text{Ge-Ge}=2.52~\mathring{A}$ and $d_%
\text{Ge-F}=1.80~\mathring{A}$. The structure is slightly puckered, with a
buckling parameter of $0.74~\mathring{A}$. The interatomic angles ranging
between $111.70%
{{}^\circ}%
$ and $111.80%
{{}^\circ}%
$ indicate an $sp^{3}$ hybridization between Ge atoms. According to Ref.~\cite{liang2016}, the Ge$_{2}$F is an antiferromagnetic semiconductor, whith a small gap energy.
\begin{figure}[tbp]
\includegraphics[scale=0.4]{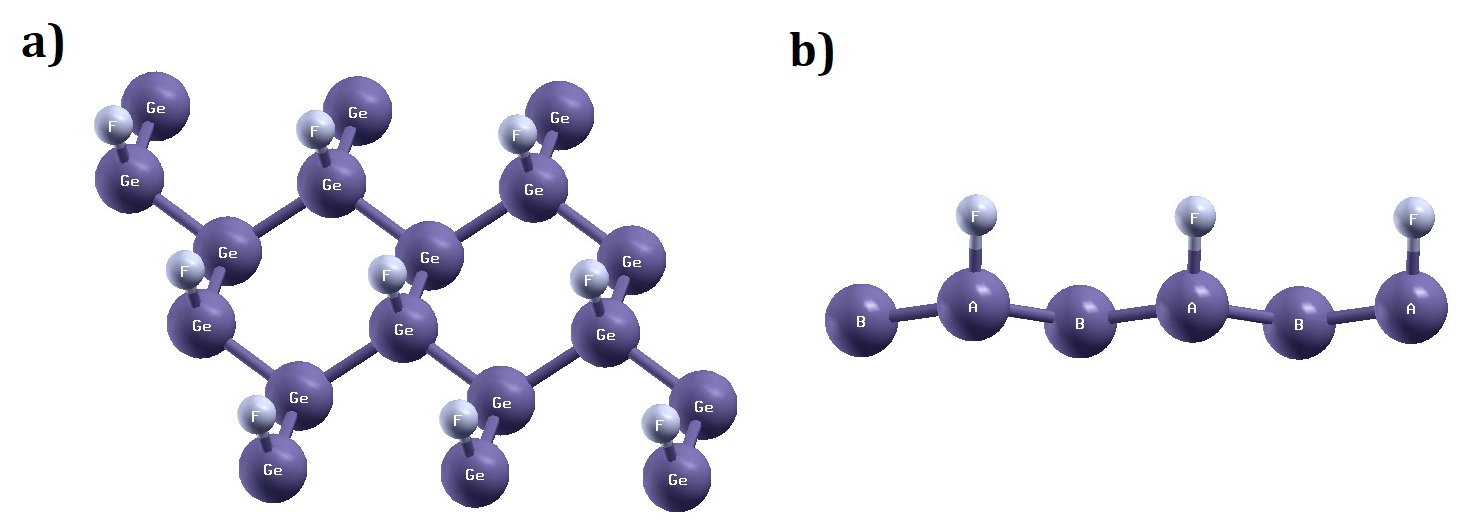} 
\caption{Optimized configuration (a) top (b) side of half-fluorinated
germanene Ge$_{2}$F. Fluor (small spheres) atoms binds to Germanium atoms (site A), while Germanium atoms at site B are undecorated (as shown in the side view configuration).}
\label{struct}
\end{figure}
To check and examine the stability of free-standing monolayer materials, various computational methods can be used such as molecular dynamics \cite{ashton,sun}, the computation of  formation energy and binding energy \cite{liang2016}, as well as translational symmetry based on the relaxation of a finite nanocluster \cite{avramov,kuklin}. In this work, the stability of half fluorinated germanene, confirmed in \cite{liang2016} calculating the formation energy, is rechecked through the analysis of  the phonon dispersion, the figure \ref{phonon} displays the phonons dispersion. Analysis of the phonon spectrum shows the absence of imaginary frequency along the high-symmetry directions of the Brillouin zone for all phonon branches. It is a signature of stability of our material.
\begin{figure}[tbp]
\begin{center}
\includegraphics[scale=0.4]{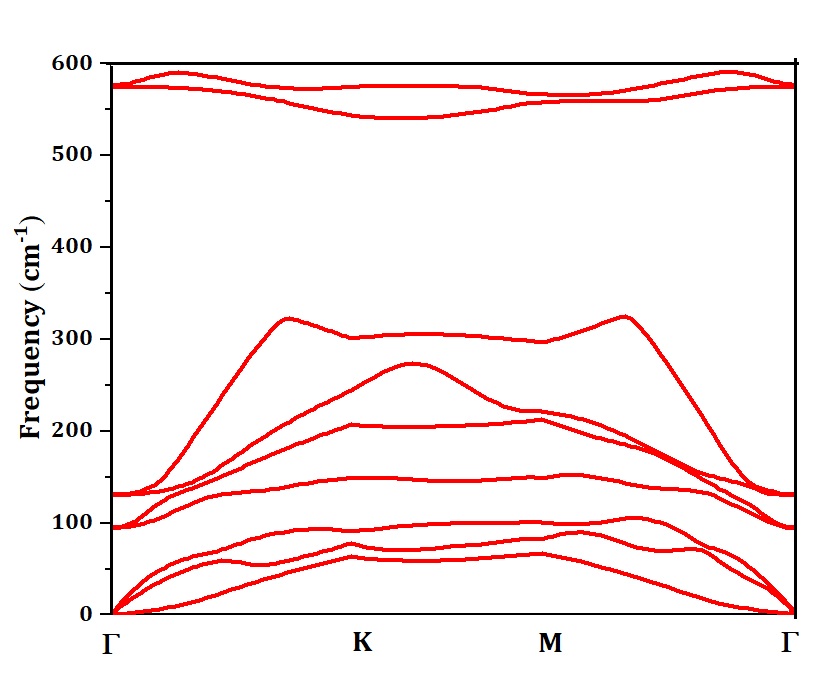} 
\caption{Phonon dispersion spectra calculated for the half fluorinated
germanene. }
\label{phonon}
\end{center}
\end{figure}
\begin{figure}[tbp]
\begin{center}
\includegraphics[scale=0.6]{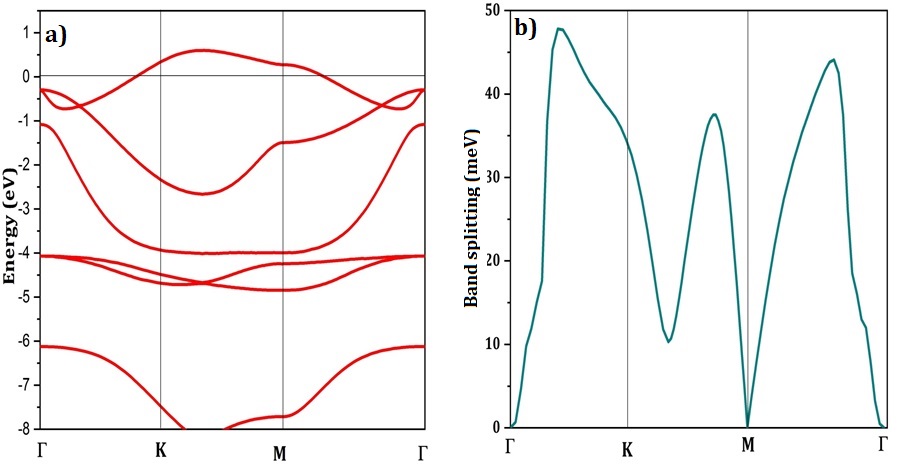}
\end{center}
\caption{ (a) The LDA band-structure with respect to the Fermi energy of
half fluorinated germanene including spin-orbit coupling. Similarly to Ref.~\protect\cite{mazurenko2016}, the impact of spin-orbit coupling on the band crossing the Fermi energy is monitored in terms of the band splitting (b).}
\label{fig1}
\end{figure}
Fig.~\ref{fig1} (a) illustrates the LDA band-structure, including SO
coupling. The metallic character of $Ge_{2}F$ is in good agreement with
recent work \cite{goli}. As expected, a small gap energy of $0.19eV$ is
reported for $Ge_{2}F$ using the generalized gradient approximation (GGA) 
\cite{liang2016}. It is worth noting that the standard DFT approximations,
namely the LDA and the GGA, are known to successfully describe the
ground-state properties and to underestimate the results of excited states.
Thus to include quasiparticle corrections, which reproduce a band gap in
accordance with the experimental measurements, one should use the GW
approximation that goes beyond the scope of this work \cite{drissi2015,
drissi2014}. Fig.~\ref{fig1} (a) also shows the bands around the Fermi level slightly overlap with other bands at the $\Gamma $-point similarly to what was found for the half fluorinated graphene (C$_{2}$F) \cite{mazurenko2016} but with a larger spin-orbit splitting induced by the heavier Ge atoms. The splitting characterizing the band crossing the Fermi energy reaches a
maximum of $48$meV, as shown in Fig.\ref{fig1} (b), which is larger than $38$meV, the maximum value reported for the half-fluorinated graphene C$_{2}$F~\cite{mazurenko2016}.
\begin{figure}[tbp]
\begin{center}
\includegraphics[width=6in]{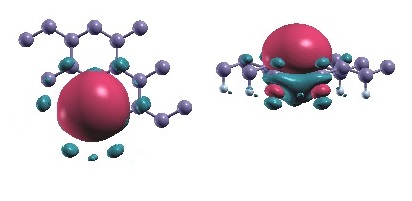}
\end{center}
\caption{Representation of Maximally-Localized Wannier functions obtained from the projection of the $p_{z}$ band at the Fermi level on the
non-decorated Ge.}
\label{figaa}
\end{figure}
For the bands located at the Fermi level, Fig.\ref{figaa} reveals that the
Wannier functions, obtained from the projection of the $p_{z} $ orbitals on the non-fluorinated Ge-atoms, are positioned at the centre of these atoms. The spread value of 2.04 indicates the delocalization of the Wannier function in real space. Besides, as shown in Fig.\ref{figaa}, the Wannier functions overlap on three nearest neighboring (NN) germanium decorated sites, giving rise to the Coulomb contribution to the total exchange
interaction. The spin up/down channel of the partial density of states, displayed in Fig.~\ref{dooo}, shows that the magnetism, which is relevant for Ge$_{2}$F near the Fermi level, is principally originated from $p_{z}$ orbitals of the
non-functionalized Ge atoms. This is due to the broken $\pi-bonding $
network of pure non magnetic germanene. More precisely, in Ge$_{2}$F, the
F-atoms form strong bonds with saturated Ge-atoms leaving $p_{z}$ electrons
of the non-decorated Ge-atoms free and localized. 
\begin{figure}[tbp]
\begin{center}
\includegraphics[scale=0.2]{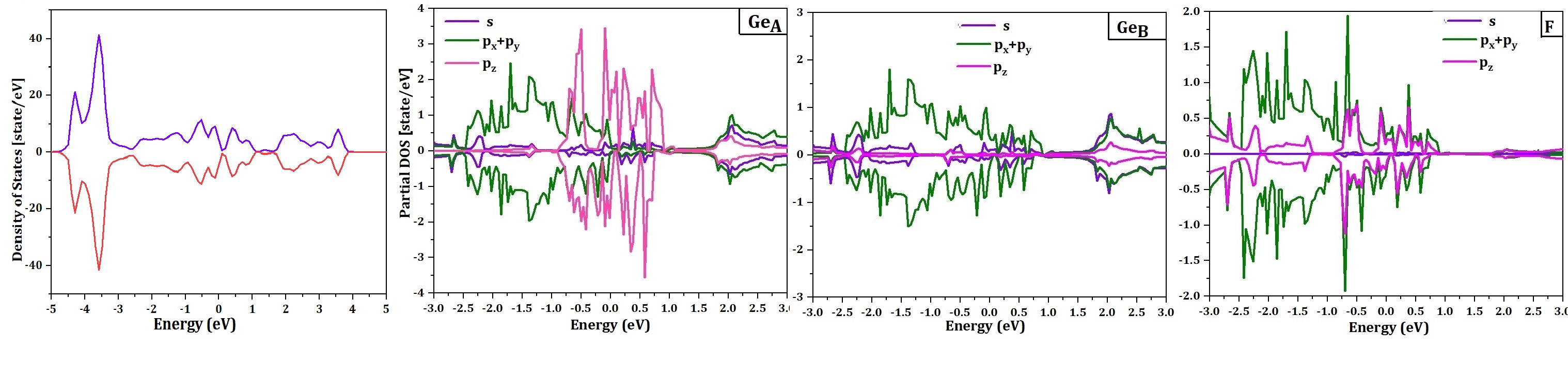}
\end{center}
\caption{Totale density of states (DOS) and Partial density of states of
half-fluorinated germanene as obtained with the antiferromagnetic (AFM)
state described in the main text and shown in Fig.~\protect\ref{mag}. The upper pannel corresponds to the majority-spin (up) channel, while the lower one hosts the minority-spin (down) channel.}
\label{dooo}
\end{figure}
Within the ab-initio formalism, we explored various magnetic configurations of Ge$_{2}$F, namely, antiferromagnetic (AFM), ferromagnetic (FM), and
ferrimagnetic (FI), utilizing the 4$\times $4 supercell shown in the Fig.~\ref{mag}. The AFM state is found to be the lowest in energy. The energy differences with respect to the non-magnetic (NM) state are: $E_\text{NM}-E_\text{AFM}=16.93$~meV, $E_\text{NM}-E_\text{FI}=15.78$~meV and $E_\text{NM}%
-E_\text{FM}=10.85$~meV.
\begin{figure}[tbp]
\begin{center}
\includegraphics[scale=0.2]{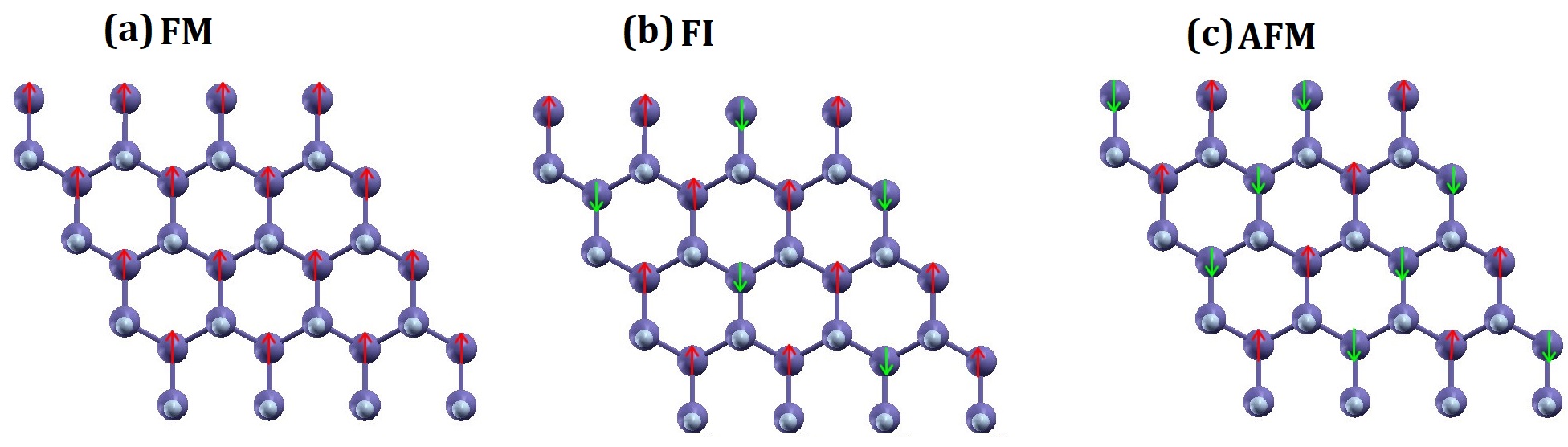}
\end{center}
\caption{Three possible magnetic configurations of half-fluorinated
germanene: (a) ferromagnetic , (b) ferrimagnetic and (c) antiferromagnetic,
as explored from ab-initio.}
\label{mag}
\end{figure}
Owing to the presence of spin-orbit coupling, the hopping integrals are $2\times2$ matrices in spin-space. They are listed below in meV for the case of first NN (01), second NN (02) and third NN (03) germanium atoms (see the schematic representation in Fig.~\ref{dzyal}(a)):
\begin{center}
\begin{equation*}
\begin{array}{cc}
t_{01}= & 
\begin{bmatrix}
-180.35+1.25i & 2.012-3.39i \\ 
-2.02-3.39i & -180.35-12.5i%
\end{bmatrix}%
\,, \\ 
t_{02}= & 
\begin{bmatrix}
11.76+0.11 & 0.51-0.88i \\ 
-0.51-0.88i & -11.758-0.11i%
\end{bmatrix}%
\,, \\ 
t_{03}= & 
\begin{bmatrix}
20.18 & -1.3+0.75i \\ 
1.3+0.75i & 20.18%
\end{bmatrix}%
\,.%
\end{array}%
\end{equation*}
\end{center}
We note that (i) the hopping matrix between the first nearest neighbors
contain large imaginary and non-diagonal elements, which are responsible for
the antisymmetric anisotropic exchange interactions (DMI), and (ii) $%
t_{ij}\ll U.$ In this case, Heisenberg Hamiltonian can be constructed within
the superexchange theory~\cite{anderson1959} as follows~\cite{yildirim1995}: 
\begin{equation}
\hat{H}=\sum_{ij}J_{ij}\hat{m}_{i}\hat{m}_{j}+\sum_{ij}\mathbf{D}_{ij}\left[ 
\hat{m}_{i}\times \hat{m}_{j}\right]  \label{eq:heisenberg}
\end{equation}%
where $\hat{m}=\frac{\mathbf{M}}{|\mathbf{M}|}$ is the classical Heisenberg
vector of unit length, $J_{ij}$ and $\mathbf{D}_{ij}$ are the isotropic
exchange coupling and the DMI vector, respectively. The summation runs twice
over all pairs.

\subsection{Isotropic exchange interaction}

A mapping of the previous Heisenberg Hamiltonian to eq.~\ref{aa} leads to
the following form of the isotropic exchange interaction: 
\begin{equation}
J_{ij}=\dfrac{1}{\tilde{U}}Tr_{\sigma }\left \{ \hat{t}_{ji}\hat{t}%
_{ij}\right \} -J_{ij}^{F}  \label{ert1}
\end{equation}%
where $\hat{t}_{ij}$ is the hopping energy taking into account spin-orbit
coupling and $\tilde{U} = U_{00}-U_{01}$\cite{mazurenko2016} corresponds to
the effective local partially-screened Coulomb interaction calculated via
constrained random phase approximation (cRPA) \cite{amadon2014}. $U_{ij}$
and $J_{ij}^{F}$, with $i \neq j$, are often much smaller than $U_{00}$,
which usually motivates their neglect. However, Mazurenko et al.~\cite%
{mazurenko2016} has shown that $J_{ij}^{F}$ needs to be taken into account
when extracting the magnetic exchange interactions in C$_2$F and C$_2$H. Our
analysis of the case of Ge$_2$F shows that in contrast to C$_2$F, the
non-local $J_{ij}^{F}$ are negligible. This can be explained by the
extremely weak spin-polarization of the non-fluorinated Ge atoms, which play
an important role in mediating the interactions between the fluorinated Ge
atoms~\cite{mazurenko2008}. The local Coulomb interaction $U_{00} $ and non
local Coulomb interaction $U_{01} $ are respectively equal to $2.80 eV $ and 
$0.92 eV $, which leads to $\tilde{U} = 1.88eV$.
The first term in equation~\ref{ert1} represents the Anderson superexchange,
while the non-local $J_{ij}^{F}$ is the ferromagnetic term that could
originate from the direct overlap of the neighboring Wannier functions. In
contrast to C$_2$F~\cite{mazurenko2016}, however, $J_{ij}^{F}$ is rather
negligible in Ge$_2$F because of the weak magnetic moment carried by
germanium. Therefore, we use in practice the usual form: \begin{equation}
J_{ij}=\dfrac{1}{\tilde{U}}Tr_{\sigma }\left \{ \hat{t}_{ji}\hat{t}%
_{ij}\right \}  \label{eq:Jij2}
\end{equation}

Using LDA+SO calculations by integrating the corresponding combination of
the Wannier functions, one deduces that the isotropic exchange interaction
between the first nearest neighbors $J_{01}=1.4$~meV is very important when
compared to the second nearest neighbors $J_{02}=0.11$~meV and the third
nearest neighbors $J_{03}=0.15$~meV. The positive sign $J_{01}$ confirms
that the half fluorinated-germanene is antiferromagnetic.

\subsection{Dzyaloshinskii-Moriya interaction}

\begin{figure}[tbp]
\begin{center}
\includegraphics[scale=0.5]{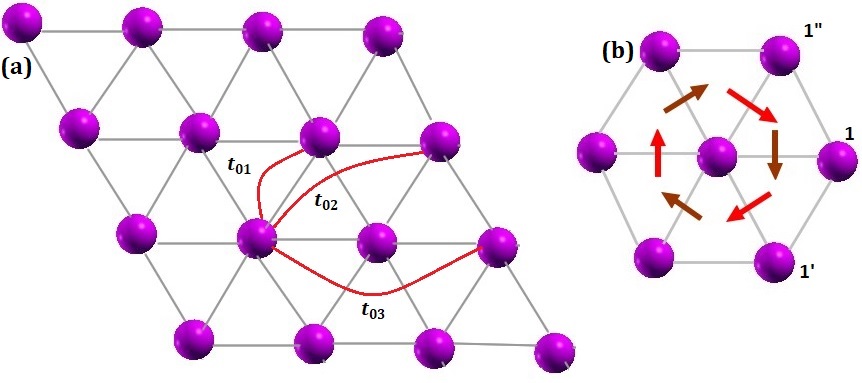}
\end{center}
\caption{{\protect\small Schematic representation of (a) Hoping paths and
(b) Dzyaloshinskii-Moriya vectors for half fluorinated germanene. Dark and
light red arrows are DM-vectors with negative and positive z-components,
respectively.}}
\label{dzyal}
\end{figure}

The anisotropic exchange interaction resulting from the spin-orbit
interaction, is expressed as follows:

\begin{equation}
\mathbf{D}_{ij}=-\dfrac{i}{2\tilde{U}}\left[ Tr_{\sigma }\left \{ \hat{t}%
_{ji}\right \} Tr_{\sigma }\left \{ \hat{t}_{ij}\sigma \right \} -Tr_{\sigma
}\left \{ \hat{t}_{ij}\right \} Tr_{\sigma }\left \{ \hat{t}_{ji}\sigma
\right \} \right]  \label{ert2}
\end{equation}%
where $\sigma $ are the Pauli matrices. For the nearest neighbour bonds in $%
G_{2}F,$ the DMI vectors as well as the radius vectors are presented in
Table~\ref{DM} and Fig.~\ref{dzyal}. 
\begin{table}[tbp]
\begin{center}
\begin{tabular}{|l|c|c|}
\hline
bond & radius vectors & $D_{ij}(meV) $ \\ \hline
$0-1^{\prime }$ & $%
\begin{pmatrix}
\dfrac{1}{2}, & -\dfrac{\sqrt{3}}{2}, & 0.0 \\ 
&  & 
\end{pmatrix}
\quad $ & $%
\begin{pmatrix}
-0.65, & -0.38, & 0.242 \\ 
&  & 
\end{pmatrix}
\quad $ \\ \hline
$0-1" $ & $%
\begin{pmatrix}
\dfrac{1}{2}, & \dfrac{\sqrt{3}}{2}, & 0.0 \\ 
&  & 
\end{pmatrix}
\quad $ & $%
\begin{pmatrix}
0.65, & -0.38, & 0.242 \\ 
&  & 
\end{pmatrix}
\quad$ \\ \hline
$0-1 $ & $%
\begin{pmatrix}
1, & 0, & 0 \\ 
&  & 
\end{pmatrix}
\quad$ & $%
\begin{pmatrix}
0.00, & -0.753, & -0.242 \\ 
&  & 
\end{pmatrix}
\quad$ \\ \hline
\end{tabular}%
\end{center}
\caption{{\protect\small The Dzyaloshinskii-Moriya vectors}}
\label{DM}
\end{table}
In general, the orientation of DMI is defined by the symmetries of the
crystal. In our case, the spin Hamiltonian symmetry is consistent with the $%
C_{3v}$ point group of the triangular lattice formed by non-functionalized
Ge atom as shown in Fig.\ref{dzyal}-b. 
The vertical reflections pass through the middle of bonds between the
nearest neighbours. Furthermore, the corresponding DMI vectors lie in the
reflection planes and are perpendicular to the interatomic bonds. The
z-components of the DMI vector can change sign depending on the pair of
nearest neighboring atoms.

\subsection{Spin-dynamics simulations}

\begin{figure}[tbp]
\caption{ (a) The ground state of half fluorinated germanene ( $Ge_ 2F$) is a N\'eel antiferromagnetic
state. The spins lay in the film plane due to the $z$-componets of the
Dzyaloshinskii-Moriya vectors ($E= -23.21$ mev/atom). (b) Applying an
external field (4000 T) along the $x$-direction (to the right-hand side), a N\'eel distorted state is obtained ($E= -38.63$ mev/atom). (c-d) Under
the influence of the same external field, metastable spin-spiral states can
be obtained ($E= -38.41$ and $E= -38.35$ mev/atom, respectively). }
\label{fig:Fig1_SW}\centering
\includegraphics[scale=1.0]{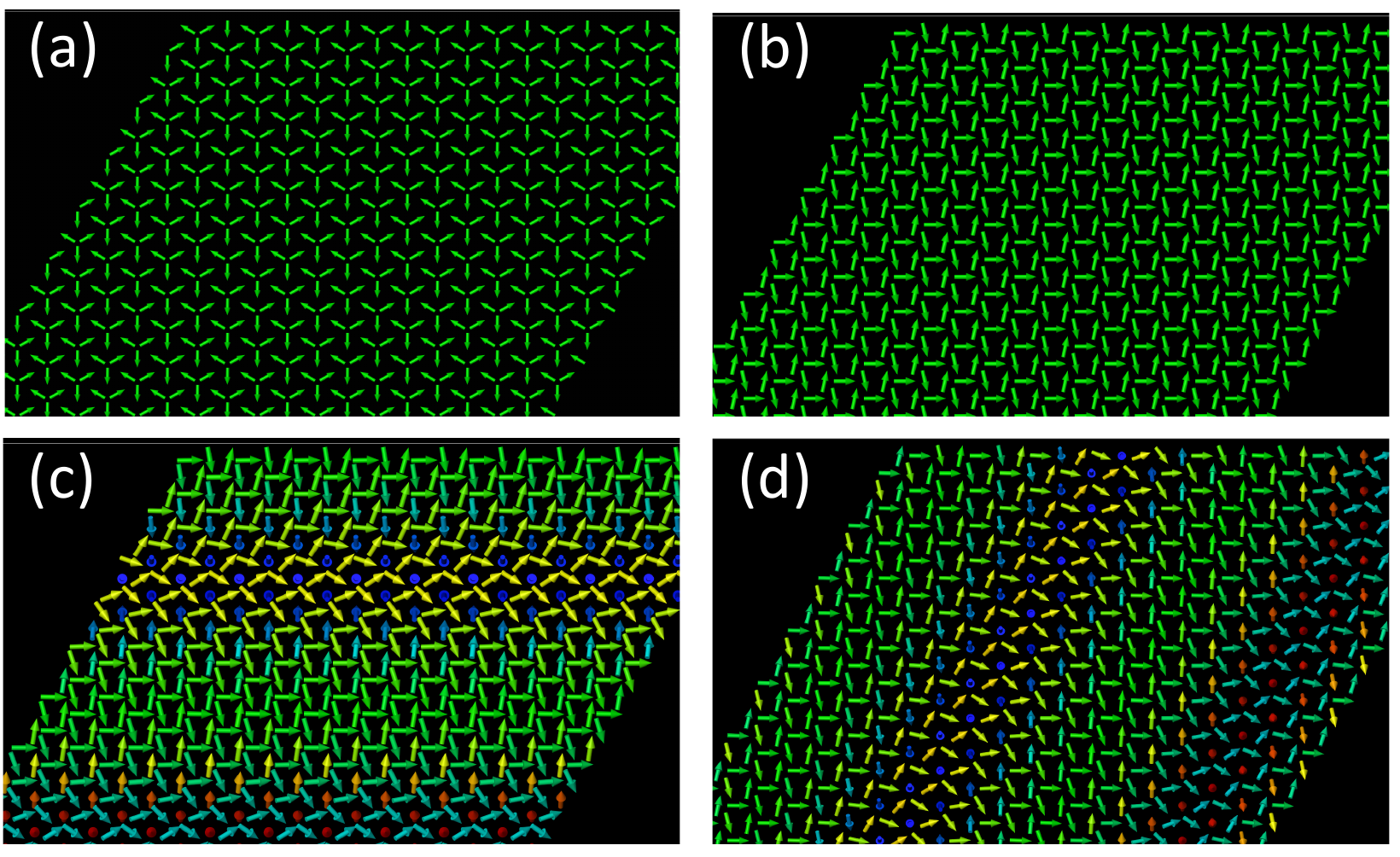}
\end{figure}

\begin{figure}[tbp]
\caption{ (a) Ground state of half fluorinated germanene ( $Ge_ 2F$) when the out-of-plane components of
the Dzyaloshinskii-Moriya vectors are disregarded (energy of -21.97
meV/atom). The spins lay in the $x-z$ plane. (b) Even in the absence of
external, spin-spiral states can be obtained (energy of -21.942 meV/atom).
(c) Applying an external field of 2000 T along the $z$-direction (to the
right-hand side), an antiferromagnetic skyrmion lattice is formed (energy of
-25.897 meV/atom). }
\label{fig:Fig2_SW}\centering
\includegraphics[width=\columnwidth]{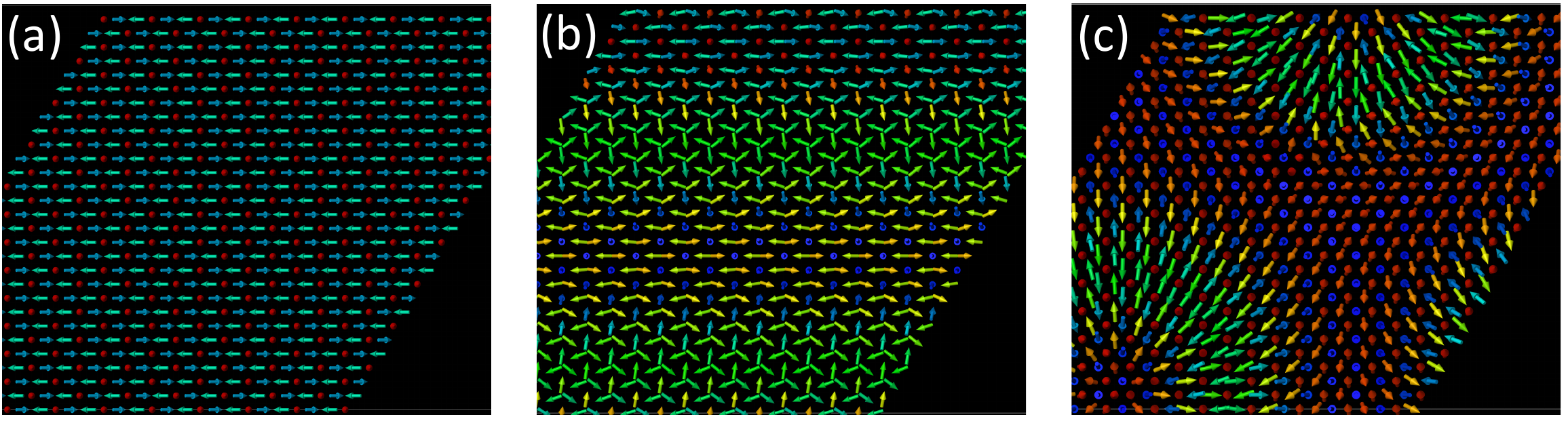}
\end{figure}

After solving the LLG equation, Eq.~\ref{LLG}, utilizing the extracted
magnetic exchange interactions, we obtained as the ground state a N\'eel
state with a zero net magnetization as shown in Fig.~\ref{fig:Fig1_SW}(a).
The nearest neighboring AFM interactions favor the realization of such a
magnetic state. Interestingly, we find that the z-component of the DMI
imposes to have the moments in the surface plane. If one removes the
z-component of the DMI, the resulting N\'eel state is characterized by
out-of-plane components of the magnetic moments (Fig.~\ref{fig:Fig2_SW}(a)).
In this particular case, a metastable spin-spiral state can be stabilized
(Fig.~\ref{fig:Fig2_SW}(b)) which has an energy of 0.031 meV/atom above that
of the ground state.

When applying a large magnetic field (up to 4000 T) along the $x$-axis,
applied to the right in Fig.~\ref{fig:Fig1_SW}(b), a modified N\'eel state
is obtained if keeping the z-component of the DMI finite. However, one can
also obtain at higher energies the spin spirals shown in Figs.~\ref%
{fig:Fig1_SW}(c--d). Without the z-component of the DMI, an
antiferromagnetic skyrmion can even be stabilized with a field of 2000 T
along the $z$-axis (Fig.~\ref{fig:Fig1_SW}(c)). We conjecture that such
large magnetic fields can be induced via a proximity effect if the 2D
material is deposited on a magnetic substrate. The equivalent magnetic
exchange energy for the 2000 T field is 32.41 meV, which could potentially be accessed.
\begin{figure}[tbp]
\begin{center}
\includegraphics[width=\columnwidth]{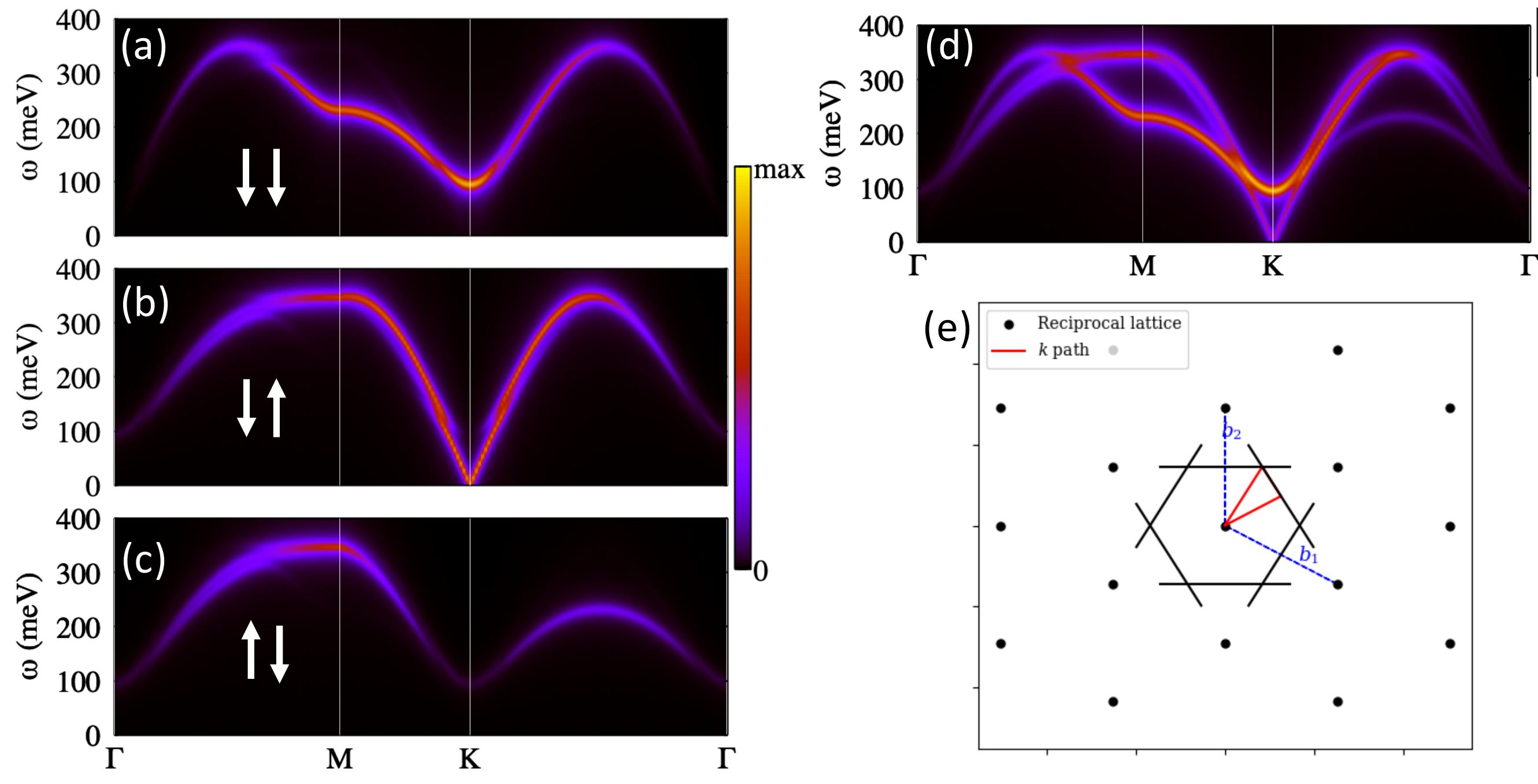}
\caption{ Inelastic electron scattering spectra for the N\'eel  ground
state. (a-c) show the spin-resolved scattering channels for probing
electrons polarized perpendicularly to the surface. (a) corresponds to a
non-spin-flip channel where we observe a linearly polarized mode (with zero
net angular momentum). The spin-wave mode observed in (b) requires a
spin-flip process. (d) displays the total spectrum that also corresponds to
a spectroscopy with unpolarized electrons. The scattering spectra was
calculated along the red path indicated in (e).}
\label{fig:Fig3_SW}
\end{center}
\end{figure}
The spin-excitations spectra corresponding to the N\'eel ground state [Fig.~%
\ref{fig:Fig1_SW}(a)] as they would be measurable with SREELS or EELS are
depicted in Figs.~\ref{fig:Fig3_SW}(a--d). We computed the spectra along the
Brillouin-zone path indicated in Fig.~\ref{fig:Fig3_SW}(e). Figure \ref%
{fig:Fig3_SW}(d) shows the total inelastic spectrum as they would be probed by an unpolarized electronic beam. Notice the strong scattering intensities around the $K$ point and the
vanishing intensities at the $\Gamma$, which are common spectroscopy
features of antiferromagnets~\cite{dosSantos:2020a,dosSantos:2020b}. We can
count up to three spin-wave branches which can be individually detected
through the spin-resolved spectroscopy. Figure~\ref{fig:Fig3_SW}(a--c)
represent the spin-resolved spectra that arise from the possible spin
orientations of the incoming and outgoing electrons. We chose the
polarization of the probing electrons perpendicular to the magnetic film.
Figure~\ref{fig:Fig3_SW}(a) corresponds to the non-spin-flip scattering
channels, such that when we send electrons with spin down we measure only
scattered electrons with the same spin. The spectrum displays a single
spin-wave branch with an energy minimum (100 meV) at the $K$ point. In Fig.~%
\ref{fig:Fig3_SW}(b) and (c), we have the spectra for the excitations which require spin flips of the probing electrons. In these processes, angular momenta is exchange between the probing electrons and the spin waves. The spin-wave mode in Fig.~\ref{fig:Fig3_SW}(b) has a linear dispersion at low energies, which are the characteristical spin-wave feature of
antiferromagnets (see for example Ref.~\cite{dosSantos:2020b}).

\section{Conclusion}
To summarize our study, we performed an ab-initio investigation of the
complex magnetic properties of a half-fluorinated Germanene (Ge$_2$F) and
use a low-energy model to map the first-principles calculations and extract
the magnetic exchange interactions as well as the Dzyaloshinskii-Moriya
interaction vector. The latter is induced by the strong spin-orbit coupling
of Germanium atoms and by the fact that the fluorine atoms break inversion
symmetry.

The magnetic exchange interactions are antiferromagnetic among the first,
second and third nearest neighbors, which stabilize a N\'eel state where the
magnetic moments are lying in the surface plane. This particular
configuration is favored by the out-of-plane component of the DMI vector.
Antiferromagnetic spin spirals are found as metastable states once a
magnetic field is applied. Interestingly, if the out-of-plane component of
the DMI vector is set to zero, antiferromagnetic skyrmions can be found. For
the realization of such chiral magnetic textures, we propose to use a
potential magnetic substrate to induce the requested large magnetic fields.
Finally, we explored for completness the spin-waves excitations and
presented the spectra that could be measurable with electron energy loss
spectrocopy or its spin-resolved version.
\section*{acknowledgement}
F. Z. Ramadan and L. B. Drissi would like to acknowledge "Acad\'{e}mie
Hassan II des Sciences et Techniques-Morocco" for financial support. L. B.
Drissi acknowledges the Alexander von Humboldt Foundation for financial
support via the George Forster Research Fellowship for experienced
scientists (Ref 3.4 - MAR - 1202992). The work of F. J. dos Santos and S.
Lounis was supported by the European Research Council (ERC) under the
European Union's Horizon 2020 research and innovation programme
(ERC-consolidator Grant No. 681405-DYNASORE). We gratefully acknowledge the
computing time granted by JARA-HPC on the supercomputer JURECA at
Forschungszentrum J\"{u}lich and by RWTH Aachen University.

\end{document}